\documentclass[twocolumn]{revtex4-2}

\usepackage[utf8]{inputenc}
\usepackage{babel}
\usepackage{fontenc}
\usepackage{graphicx}
\usepackage{amsmath}
\usepackage{subcaption}
\usepackage{verbatim}
\usepackage{color}
\usepackage{amssymb}
\usepackage{mathrsfs}
\usepackage{comment}

\usepackage[normalem]{ulem}

\begin{document}

\author{Dalibor Jav\r{u}rek}
\email{dalibor.javurek@mojeposta.xyz}

\date{\today}

\title{Extension of Theory of Gravitomagnetism and Spinor Quantum Mechanics with Dynamics of Free Electromagnetic Field}

\begin{abstract}
The theory of Gravitomagnetism and spinor quantum mechanics describing the interaction between the Dirac spinor field, the electromagnetic field, and a weak gravitational field is extended by including the Lagrangian density of the free electromagnetic field. It is shown that the newly added term in the Lagrangian density is necessary to restore a symmetric energy-momentum tensor in the interaction term of the Lagrangian density within the theory. We prove that when the electromagnetic energy term is missing, the tensor in the interaction Lagrangian density does not represent the energy-momentum tensor of the Dirac field minimally coupled to the electromagnetic field on flat space-time. Moreover, it follows that the Lagrangian density remains unchanged regardless of whether the theory is developed from the principle of general covariance or is defined as a flat space-time theory with the gravitational field treated as a regular tensor field, in analogy to the electromagnetic field. This contradicts previous findings.
\end{abstract}

\maketitle

\section{Introduction}
In the exploration of classical systems within external gravitational fields, extensive attention has been devoted to a variety of gravitational effects. Recent investigations have successfully affirmed the presence of the gravitomagnetic, or frame-dragging, field induced by the Earth's rotation. This phenomenon was conclusively observed in two distinct experiments, involving LAGEOS satellites~\cite{Ciufolini1997,Ciufolini1995} and the Gravity Probe B (GP-B) satellite~\cite{Fraser2006,Will1993,Will2018,Everitt2011}. LAGEOS provided a measurement of the gravitomagnetic interaction through its influence on satellite orbits, while the GP-B satellite validated the general relativity prediction for the gravitomagnetic precession of a gyroscope in Earth's orbit.
The LAGEOS effects originated from Lorentz-like forces in the geodesic equation, whereas the GP-B effects were attributed to Larmor-like torques from the spin precession equation. These distinctions rendered the two experiments independent and mutually complementary \cite{Adler2012}.

In the realm of quantum systems interacting with gravitational fields, notable experiments and theoretical studies have been conducted in relation to Earth's gravitational influence. These investigations include studies of neutrons interacting with the Earth's Newtonian field~\cite{Nesvizhevsky2002,Tobar2022} as well as atom interferometer experiments aimed at rigorously testing the equivalence principle~\cite{Zych2015}. Moreover, a gravitational wave detector composed of two separated atom interferometers has been proposed~\cite{Dimopoulos2008}. Although attempts to discern gravitomagnetic effects with these devices have been discussed, the formidable challenges arising from the effects' small scale and their similarity to classical rotational effects make such experiments inherently difficult~\cite{Adler2012}.

The Dirac field in curved space-time represents a key area of research that connects quantum mechanics, quantum field theory, and general relativity. Its importance lies in its applications in understanding the universe, particularly in astrophysics and cosmology. Additionally, it allows for the discovery of novel types of effects following from the interaction of curved space-time, the quantum-mechanical Dirac field, and the electromagnetic field. Hopefully, the research in this area will indicate whether the wave-function of a quantum system is much more closely related to the gravitational interaction than previously expected~\cite{Maldacena2013,Susskind2016}.

Recent research on Dirac fields in curved space-time has expanded our knowledge in several ways. A rigorous while still pedagogically valuable study has looked into how quantized Dirac fields behave in curved space-times with low relativistic effects \cite{Falcone2022}. The influence of two-dimensional space curvature on massless Dirac particles has been examined \cite{Flouris2018}, and the criteria for quantizing Dirac fields in curved space-times have been set \cite{Cortez2020}. Attention has been devoted to the exploration of quantum walks related to massless Dirac fermions in curved space-time \cite{Dimolfetta2013} and to entropic dynamics \cite{Ipek2019}. Detailed analyses of solutions and oscillator models in curved environments have been performed \cite{Oliveira2019, Oliveira2020,Yagdjian2021}. The research demonstrates the ongoing effort to understand how quantum fields interact with the structure of space-time.

Many other interesting effects resulting from the gravitational field coupling to the Dirac field have been studied. Attention has been devoted to the investigation of parity violation due to the Dirac field's gravity-spin interaction~\cite{Jentschura2013}. The interplay between Einstein's equivalence principle, atomic transition frequencies, and the g-factor measurements has been explored~\cite{Jentschura2018}. The symmetry guaranteeing the same gravitational attraction for electrons and positrons has been found~\cite{Jentschura2013b}. Despite all the knowledge already obtained about the topic, we believe there are still remarkable questions and areas yet unexplored.

The paper analyzes and extends the theory of Gravitomagnetism and spinor quantum mechanics (GQM)~\cite{Adler2012}, which covers the interaction of the Dirac field, the electro-magnetic field and the weak gravitational field. All fields in the theory are assumed to be classical. We find the original GQM theory remarkable and highly valuable. Its value lies in the development of the theory from principles of general covariance and an effort to analyse, if the theory can be transformed into a flat space-time theory. This could be highly beneficial for physicists already familiar with flat space-time quantum field theory.  In the flat space-time theory, the gravitational field tensor describing disturbance in the metric is treated as an ordinary tensor; in analogy to the electro-magnetic field. It has been concluded~\cite{Adler2012}, that when the theory is developed from principles of general covariance, purely geometric terms related to the metric appear in the total Lagrangian density. It follows, the theory cannot be described by a Lagrangian density of a flat space-time theory in a physically satisfactory way. In this paper, we show, that the Lagrangian density of the Gravitomagnetic theory can be recast into the Lagrangian density of a physically meaningful flat space-time theory if the original theory is extended by a term related to free electro-magnetic field. We consider this finding to be significant, since it may allow the theory to be more accessible to physicists already acquainted with flat space-time theories. Moreover, we show that the extension is necessary in order to obtain symmetric energy-momentum tensor in the interaction term of the theory.

After a brief introduction to the topic has been given in Sec.~I, we summarize the GQM theory, as it has been developed by Adler~et.~al. in Sec.~II. In Sec.~III together with Appendix~A, we derive and analyze the energy-momentum tensor of the corresponding flat space-time theory, which finds its place in the GQM theory. This is essential for the physical interpretation of the Lagrangian density of the theory and for the definition of the Gravitomagnetic theory as a flat space-time theory. In Sec.~IV, the Lagrangian density of the Gravitomagnetic theory is extended with a term related to the free electro-magnetic field. It is shown, that the addition of this term is necessary in order to obtain a symmetric energy-momentum tensor in the interaction Lagrangian density. In Sec.~V, the equivalence between the definition of the theory from the principle of general covariance and definition as a flat space-time theory with the gravitation field treated as a tensor field is shown. Sec.~VI summarizes all findings mentioned in the paper. Appendix~A is devoted to the derivation of the flat space-time energy-momentum tensor related to the non-extended Gravitomagnetic theory.

\section{Theory of Gravitomagnetism and Spinor Quantum Mechanics}
A theory describing interaction between the electro-magnetic four-potential field $A^\mu$ and the Dirac spinor field $\psi$ on weakly curved space-time has been developed by Adler~et.~al. \cite{Adler2012}. Since the gravitational field has been assumed weak, coefficients of the metric tensor $g_{\mu \nu}$ are defined by equation
\begin{equation}
\label{e01}
g_{\mu \nu} = \eta_{\mu \nu} + h_{\mu \nu}. 
\end{equation}
$\eta_{\mu \nu} = \mathrm{diag}(1,-1,-1,-1)$ is the metric tensor of the flat space-time and the coefficients of the tensor $h_{\mu \nu}$ describe small perturbations on otherwise flat space-time. The coefficients $h_{\mu \nu}$ are assumed to be much smaller than one $ \vert h_{\mu \nu} \vert \ll 1$ in absolute value. The matrix of the metric disturbance tensor $h_{\mu \nu}$ is given by the equation
\begin{equation}
\label{e02}
 h_{\mu \nu} =
 \left[
 \begin{array}{cccc}
 2\phi & h^1 & h^2 & h^3 \\
 h^1 & 2 \phi & 0 & 0  \\
 h^2 & 0 & 2 \phi & 0  \\
 h^3 & 0 & 0 & 2 \phi
 \end{array}
 \right],
\end{equation}
where $\phi$ denotes the scalar gravitational potential and $\mathbf{h}~=~[h^1, h^2, h^3]$ is the gravitational vector potential. Geometrized units $c = \hbar = \mu_0 = \varepsilon_0 = 1$ are used in the paper. We remark, that this choice of $h_{\mu \nu}$ tensor results in the geodesic equation (for small speeds and terms linear in $h_{\mu\nu}$)
\begin{equation}
 \dfrac{d^2 x^i}{d t^2} = a^i \approx - \Gamma^i_{00} = \partial_t h^i - \partial_i \phi.
\end{equation}
We further note, that the theory assumes, that the magnetic field changes slowly with time , thus $\partial_t h^i = 0$.

The theory~\cite{Adler2012}  has been developed with the principle of general covariance. Particularly, the GQM theory has been based upon the generalization of the Lagrangian density of the Dirac field $\psi$ minimally coupled to the electro-magnetic field $A^\mu$ on flat space-time to a general space-time by the condition, that the gravitational field associated with the curved space-time is weak. It has also been assumed, that the sources of the gravitational field are moving slowly. After simplification and rearranging of terms in the Lagrangian density of the theory $L$, it has been shown that the Lagrangian density $L$ attains a form
\begin{align}
\label{e1}
 L &= L_{\rm flat} + L_{\rm IG}\\
 \nonumber
 L_{\rm flat} &=  \dfrac{i}{2} [\bar{\psi} \gamma^\mu (\partial_\mu \psi) - (\partial_\mu \bar{\psi}) \gamma^\mu \psi]\\
\label{e2}
&- m \bar{\psi} \psi - q A_\mu \bar{\psi} \gamma^\mu \psi \\
\nonumber
L_{\rm IG} &= -\dfrac{h_{\mu \nu}}{4} \left[ \bar{\psi} \hat{\gamma}^\nu(i \partial^\mu \psi - q A^\mu \psi) - (i \partial^\mu \bar{\psi} + qA^\mu \bar{\psi}) \hat{\gamma}^\nu \psi \right] \\
\label{e3}
&= -\dfrac{1}{2} h_{\mu \nu} T^{\mu \nu};
\end{align}
\begin{align}
\nonumber
T^{\mu \nu}&= \dfrac{i}{4}(\bar{\psi} \gamma^\mu \partial^\nu \psi - \partial^\nu \bar{\psi} \gamma^\mu \psi + \bar{\psi} \gamma^\nu \partial^\mu \psi - \partial^\mu \bar{\psi} \gamma^\nu \psi) + \\
\label{e31}
&-\dfrac{q}{2}(\bar{\psi} A^\nu \gamma^\mu \psi + \bar{\psi} A^\mu \gamma^\nu \psi).
\end{align}
The $L_{\rm flat}$ is the Lagrangian density of the Dirac field $\psi$ minimally coupled to the electro-magnetic field $A^\mu$ on flat space-time, $L_{\rm IG}$ is the interaction Lagrangian density describing the interaction of the gravitational field $h_{\mu \nu}$ with the electro-magnetic field $A^\mu$ and the Dirac field $\psi$; $\gamma^\mu$ are flat space-time Dirac matrices obeying the anti-commutation relation
\begin{equation}
[\gamma^\mu,\gamma^\mu]_+ = \gamma^\mu \gamma^\nu + \gamma^\nu \gamma^\mu= 2 \eta^{\mu \nu}. 
\end{equation}
The field $\bar{\psi}$ is related to the Dirac field $\psi$ through the relation $\bar{\psi} = \psi^\dagger \gamma^0$.  It has been shown~\cite{Adler2012}, that the interaction Lagrangian density $L_{\rm IG}$ can be written as a scalar product between the gravitational field tensor $h_{\mu \nu}$ and the tensor $T^{\mu \nu}$; in the same manner as the interaction Lagrangian between the electro-magnetic field $A^\mu$ and the electric four-current $J^\mu = q\bar{\psi} \gamma^\mu \psi$. The tensor $T^{\mu \nu}$ has been claimed to be the energy-momentum tensor of the Dirac field minimally coupled to the electro-magnetic field $A^\mu$ on flat space-time~\cite{Adler2012}.

The action of the theory $S$ is given by a volume integral over the whole space-time, which is considered to be weakly curved
\begin{equation}
\label{e32}
 S = \int d^4x \sqrt{-g}\, L.
\end{equation}
In Eq.~(\ref{e32}) $g$ denotes the determinant of the metric tensor $g_{\mu \nu}$. In accordance with Adler~et.~al., let us denote 
the overall Lagrangian density under the integral in Eq.~(\ref{e32}) as
\begin{equation}
 \mathcal{L} = \sqrt{-g}\, L.
\end{equation}
Physically, the quantity $\mathcal{L}$ can be considered the Lagrangian density of a theory on flat space-time $x^\mu$. However, it has to be kept in mind that the coordinates $x^\mu$ are not physical. 

\section{Energy-Momentum Tensor of the Non-Extended Theory}
Knowledge of the energy-momentum tensor of a theory describing the Dirac field minimally coupled to the electro-magnetic field on flat space-time is essential, because of its identification in the Lagrangian density of the GQM theory $L_{\rm IG}$, see Eq.~(\ref{e3}). Moreover, the tensor appearing in the interaction Lagrangian density of the GQM theory is required to be a symmetric energy-momentum tensor, since it could be considered as a source of the tensor gravitational field $h_{\mu \nu}$~\footnote{if the Lagrangian density of the free weak gravitational field $h_{\mu \nu}$ were included into the overall Lagrangian density $L$}, in accordance with the Einstein equations of the general theory of relativity. In this section, we prove that the energy-momentum tensor $T^{\mu \nu}$ of the GQM theory, see Eq.~(\ref{e31}), does not satisfy the required relation with the canonical energy-momentum $\Theta^{\mu \nu}$. Therefore, it is not the energy-momentum tensor of the corresponding flat space-time theory. We show, that the energy-momentum tensor derived with the second Noether theorem satisfying this relation is distinct and asymmetric. We comment on the asymmetric nature of the tensor.  

In paper~\cite{Adler2012}, where the GQM theory has been developed, the authors derived energy-momentum tensor $T^{\mu \nu}$ of the Dirac field minimally coupled to the electro-magnetic field $A^\mu$, see appendix~B in~\cite{Adler2012}. They started with the derivation of the canonical energy-momentum tensor for the free Dirac field on flat space-time. Then, they correctly stated, that the canonical energy-momentum tensor can be symmetrized into the form in Eq.~(B3) in \cite{Adler2012}. 
To include the electro-magnetic interaction into the energy-momentum tensor, they exchanged all the derivatives multiplied by an imaginary unit $i\partial_\mu$ in the tensor for $i\partial_\mu - q A_\mu$. They arrived at a symmetric energy-momentum tensor $T^{\mu \nu}$ in Eq.~(\ref{e31}), which can be rewritten into the form
\begin{align}
\label{e4}
T^{\mu \nu} &= \dfrac{1}{2}\left( \Theta^{\mu \nu} + \Theta^{\nu \mu} -  q A^\mu \bar{\psi} \gamma^\nu \psi - q A^\nu \bar{\psi} \gamma^\mu \psi \right);\\
\Theta^{\mu \nu} &= \dfrac{i}{2} \left[ \bar{\psi} \gamma^\mu (\partial^\nu \psi) -  (\partial^\nu \bar{\psi}) \gamma^\mu \psi \right],
\end{align}
where $\Theta^{\mu \nu}$ denotes the canonical energy-momentum tensor of the free Dirac field, which is not symmetric. The validity of the energy-momentum tensor in Eq.~(\ref{e4}) was justified by the calculation of its divergence
\begin{equation}
\label{e5}
\partial_\mu T^{\mu \nu} = -q(\bar{\psi} \gamma_{\mu} \psi) F^{\nu \mu} = - F^{\nu \mu} J_\mu.
\end{equation}
where $F^{\nu \mu}$ is the electro-magnetic field tensor on flat space-time. 

We derive the energy-momentum tensor $T^{\mu \nu}_{\rm N}$ from the Lagrangian density $L_{\rm flat}$ in Eq.~(\ref{e2}) with the second Noether theorem~\cite{Freese2022}. We utilize the theorem, since it is considered to provide symmetric energy-momentum tensors~\cite{Freese2022}. The derived energy-momentum tensor $T^{\mu \nu}_{\rm N}$ is given by the equation
\begin{equation}
\label{e6}
T^{\mu \nu}_{\rm N} = \dfrac{1}{2}(\Theta^{\mu \nu} + \Theta^{\nu \mu} + q \bar{\psi} A^\nu \gamma^\mu \psi - q \bar{\psi} A^\mu \gamma^\nu \psi).
\end{equation}
For details about the derivation of the tensor, please, see Appendix~A. The tensor $T^{\mu \nu}_{\rm N}$ in Eq.~(\ref{e6}) consists of a symmetric part $\Theta^{\mu \nu} + \Theta^{\nu \mu}$ and a remaining anti-symmetric part, explicitly dependent on components of the electro-magnetic field $A^\mu$. Therefore, the obtained overall energy-momentum tensor $T^{\mu \nu}_{\rm N}$ is not symmetric and not equal to the energy-momentum tensor $T^{\mu \nu}$ obtained in Eq.~(\ref{e4}). The asymmetry in the tensor $T^{\mu \nu}_{\rm N}$ occurs, because the electro-magnetic field $A^\mu$ transfers its angular momentum to the Dirac field, but its intrinsic angular momentum is not accounted for in the tensor $T^{\mu \nu}_{\rm N}$. This occurs, since the Lagrangian density $L_{\rm flat}$ in Eq.~(\ref{e2}) lacks a term related to the free electro-magnetic field. In this work, we define the electro-magnetic field tensor $F^{\mu \nu}$ on flat space-time by the equation
\begin{equation}
 F^{\mu \nu} = \partial^\mu A^\nu - \partial^\nu A^\mu.
\end{equation}

There are two distinct energy momentum tensors calculated for the theory defined by the Lagrangian density $L_{\rm flat}$ in Eq.~(\ref{e2}) . The symmetric one $T^{\mu \nu}$ in Eq.~(\ref{e4}), derived by the authors in~\cite{Adler2012}, and an asymmetric one $T^{\mu \nu}_{\rm N}$ in Eq.~(\ref{e6}). We prove, that the energy-momentum tensor $T^{\mu \nu}$ in Eq.~(\ref{e4}) is not the energy-momentum tensor related to the flat space-time theory described by the Lagrangian density $L_{\rm flat}$ in Eq.~(\ref{e2}). This follows from the expression for the four-divergence of the tensor $T^{\mu \nu}$ in Eq.~(\ref{e5}). It has to be equal to four-divergence of canonical energy-momentum tensor $\Theta^{\mu \nu}$ (see Eq.~(\ref{ea14})), which it is not
\begin{equation}
\label{e7}
 \partial_\mu T^{\mu \nu} \neq \partial_\mu \Theta^{\mu \nu} = - \dfrac{\partial L_{\rm flat}}{\partial A_\mu} (\partial^\nu A_\mu).
\end{equation}
For details, please, see Appendix~A. On the other hand, the four-divergence of the energy-momentum tensor $T^{\mu \nu}_{\rm N}$ satisfies the first equality in Eq.~(\ref{e7}).

In consequence, the tensor $T^{\mu \nu}$ in the interaction Lagrangian $L_{\rm IG}$ in Eq.~(\ref{e3}) is not the energy-momentum tensor related to the flat space-time theory. On the other hand, the tensor $T^{\mu \nu}_{\rm N}$ is the energy-momentum tensor of the theory, since its four-divergence is equal to the four-divergence of the canonical energy-momentum tensor $\Theta^{\mu \nu}$. As an energy-momentum tensor, it can be written as an extension of the canonical momentum tensor $\Theta^{\mu \nu}$ with the four-divergence of the Belinfante tensor $B^{\lambda \mu \nu}$, see Eq.~(\ref{ea5}). Thus, the asymmetric tensor $T^{\mu \nu}_{\rm N}$ is the energy-momentum tensor related to the flat space-time theory defined by Lagrangian density $L_{\rm flat}$ in Eq.~(\ref{e2}). A symmetric energy-momentum tensor in the interaction term of the Lagrangian density $L_{\rm IG}$ appears if the theory is extended with the dynamics of the free electro-magnetic field.

\section{Extension of the Theory with Lagrangian Density of Free Electromagnetic Field}

To obtain a symmetric energy-momentum tensor $T^{\mu \nu}_{\rm N}$  with the second Noether theorem from a flat space-time theory based on the Lagrangian density $L_{\rm flat}$ in Eq.~(\ref{e2}), it is necessary to extend the Lagrangian density with Lagrangian density of free electro-magnetic field $A^\mu$. Then, we obtain Lagrangian density of quantum electrodynamics on flat space-time~\cite{Freese2022} 
\begin{equation}
\label{e8}
 L_{\rm QED, flat} = L_{\rm flat} - \dfrac{1}{4} F^{\mu \nu} F_{\mu \nu}.
\end{equation}
The energy-momentum tensor derived from the Lagrangian density in Eq.~(\ref{e8}) with the second Noether theorem is equal to~\cite{Freese2022}
\begin{align}
\nonumber
T^{\mu \nu}_{\rm QED} &= T^{\mu \nu}_{\rm N} + F^{\mu}_{\phantom{\mu} \lambda} F^{\lambda \nu} + \dfrac{\eta^{\mu \nu}}{4}F^{\rho \lambda} F_{\rho \lambda} - q A^\nu \bar{\psi}
\gamma^\mu \psi \\
\nonumber
&= \Theta^{\mu \nu} + \Theta^{\nu \mu} - \dfrac{q}{2}( A^\nu \bar{\psi} \gamma^\mu \psi + A^\mu \bar{\psi} \gamma^\nu \psi) +  F^{\mu}_{\phantom{\mu} \lambda} F^{\lambda \nu}\\
\label{e9}
&- \eta^{\mu \nu} L_{\rm QED, flat}.
\end{align}
We note, that $L_{\rm flat} \equiv 0$ due to the Euler-Lagrange equations (Dirac equations) for $\psi$ and $\bar{\psi}$. Therefore in Eq.~(\ref{e9}), we may put the Lagrangian density of quantum electrodynamics $L_{\rm QED, flat}$ equal to the Lagrangian density of the free electro-magnetic field
\begin{equation}
 L_{\rm QED, flat} = -\dfrac{1}{4} F^{\mu \nu} F_{\mu \nu}.
\end{equation}

The tensor $T^{\mu \nu}$ in Eq.~(\ref{e4}) of the GQM theory, obtained in the interaction Lagrangian density $L_{\rm IG}$ in Eq.~(\ref{e3}), constitutes a part of the energy momentum tensor $T^{\mu \nu}_{\rm QED}$ in Eq.~(\ref{e9}). Particularly,
\begin{equation}
\label{e10}
 T^{\mu \nu}_{\rm QED} = T^{\mu \nu} +  F^{\mu}_{\phantom{\mu} \lambda} F^{\lambda \nu} - \eta^{\mu \nu}L_{\rm QED, flat}.
\end{equation}
Therefore, it is tempting to extend the Lagrangian density of the GQM theory $L$ in Eq.~(\ref{e1}) with the Lagrangian density of the free electro-magnetic field on weakly curved space-time
\begin{equation}
\label{e11}
L_{\rm EM} = -\dfrac{1}{4} F^{\mu \nu} F_{\mu \nu} = -\dfrac{1}{4} g^{\mu \lambda} g^{\nu \rho} F_{\lambda \rho} F_{\mu \nu}
\end{equation}
and see, if the energy-momentum tensor $T^{\mu \nu}_{\rm QED}$ emerges in the interaction Lagrangian density $L_{\rm IG}$ instead of the tensor $T^{\mu \nu}$.

The contra-variant metric tensor $g^{\mu \nu}$ has approximately the form
\begin{equation}
\label{e12}
 g^{\mu \nu} \approx \eta^{\mu \nu} - \eta^{\mu \rho} \eta^{\nu \lambda} h_{\rho \lambda} \approx \eta^{\mu \nu} - h^{\mu \nu}.
\end{equation}
In the approximation, the second and higher order terms in the gravitational field tensor  components $h_{\mu \nu}$ are neglected. The tensor of the electro-magnetic field $F_{\mu \nu}$ on a curved space-time is defined with covariant derivatives $\nabla_\mu$ by the equation~\cite{Bunney2022}
\begin{equation}
\label{e13}
 F_{\mu \nu} = \nabla_\mu A_\nu - \nabla_\nu A_\mu.
\end{equation}
But due to the symmetry of the Christoffel symbols $\Gamma^{\nu}_{\mu \rho}~=~\Gamma^\nu_{\rho \mu}$~\cite{Bunney2022}, the electro-magnetic field tensor $F_{\mu \nu}$ attains the same form as on flat-space-time
\begin{equation}
\label{e14}
 F_{\mu \nu} = \nabla_\mu A_\nu - \nabla_\nu A_\mu = \partial_\mu A_\nu - \partial_\nu A_\mu.
\end{equation}

If we substitute for the metric tensor $g^{\mu \nu}$ from Eq.~(\ref{e12}) into Eq.~(\ref{e11}) and use the anti-symmetric property of the electro-magnetic field tensor $F_{\mu \nu} = -F_{\nu \mu}$ and  the symmetric property of the gravitational field tensor $h_{\mu \nu} = h_{\nu \mu}$, we obtain the Lagrangian density of the electro-magnetic field $L_{\rm EM}$ up to the first order in $h_{\mu \nu}$
\begin{equation}
\label{e15}
 L_{\rm EM} \approx -\dfrac{1}{4} \eta^{\mu \rho} \eta^{\nu \lambda} F_{\rho \lambda} F_{\mu \nu} - \dfrac{1}{2} F^{\mu \lambda} F_{\lambda}^{\phantom{\lambda} \nu} h_{\mu \nu}.
\end{equation}
The first term is equal to the Lagrangian density of the electro-magnetic field on flat space-time, due to Eq.~(\ref{e14}). The second term in Eq.~(\ref{e15}) describes the interaction of the electro-magnetic field $A^\mu$ with the gravitational tensor field $h_{\mu \nu}$.

Since the Lagrangian density $L_{\rm EM}$ is added to Lagrangian density of the GQM theory $L$ (see Eq.~(\ref{e1})), we transfer the second term in Eq.~(\ref{e15}) into the interaction Lagrangian density $L_{\rm IG}$, see Eq.~(\ref{e3}), where it reconstructs part of the energy-momentum tensor $T^{\mu \nu}_{\rm QED}$ in Eq.~(\ref{e10}) related to the free electro-magnetic field. As a result, we can write the Lagrangian density of quantum electrodynamics on a weakly curved space-time $L$ as the sum of the Lagrangian density of the same theory on a flat space-time and an interaction term related to the tensor gravitational field $h_{\mu \nu}$
\begin{equation}
\label{e16}
 L = L_{\rm QED, flat} - \dfrac{1}{2}h_{\mu \nu}(T^{\mu \nu} + F^{\mu \lambda} F_{\lambda}^{\phantom{\lambda} \nu}). 
\end{equation}
The tensor, which multiplies the gravitational field tensor $h_{\mu \nu}$ in the second term of Eq.~(\ref{e16}) is missing the diagonal term $-\eta^{\mu \nu} L_{\rm QED, flat}$ in order to fully reconstruct the energy-momentum tensor of quantum electrodynamics $T^{\mu \nu}_{\rm QED}$ in Eq.~(\ref{e10}).

In order to rectify this situation, it is necessary to work with the Lagrangian density 
\begin{equation}
\label{e17}
\mathcal{L} = L \sqrt{-g},
\end{equation}
which appears under the four-dimensional volume integral for the action $S$ in Eq.~(\ref{e32}).  The Lagrangian density $\mathcal{L}$ can be considered as defining a flat space-time theory on coordinates $x^\mu$, which are not physical. 

\section{Equivalence of Gravitomagnetic Theory with Flat Space-Time Theory}
It has been noted~\cite{Adler2012}, that the GQM theory posses purely geometric term in its total Lagrangian density $\mathcal{L}$ in Eq.~(\ref{e17}) when developed from principles of general covariance. Thus, if the weak gravitational field $h_{\mu \nu}$ was treated as a tensor field propagating on a flat space-time coupled to other fields on flat space-time in the definition of this theory, some terms in the resulting Euler-Lagrange equation for the Dirac field $\psi$ would be missing. On the other hand, Feynman, Weinberg and Schwinger have all been pointing out, that the geometric nature of the gravitational field does not have to be taken into account~\cite{Adler2012,Weinberg1972,Feynman2002} and can be treated as an ordinary field on flat space-time as far as the gravitational field is weak. Here, we show the equivalence between the GQM theory extended with a free electro-magnetic term (see SEC~III.) and an idea proposed by Feynman, Weinberg and Schwinger.

We start with calculation of term $\sqrt{-g}$. Up to the first order in components of gravitational field tensor $h_{\mu \nu}$. It is equal to~\cite{Adler2012}
\begin{equation}
\label{e18}
 \sqrt{-g} \approx 1 + \dfrac{1}{2}h^\mu_{\phantom{\mu} \mu} \approx 1 + \dfrac{1}{2}\eta^{\mu \nu} h_{\mu \nu}
\end{equation}
The Lagrangian density $\mathcal{L}$ in Eq.~(\ref{e17}) can be written up to the first order in terms of gravitational field tensor $h_{\mu \nu}$ as
\begin{align}
\label{e19}
 \mathcal{L} \approx L_{\rm QED, flat} \sqrt{-g} + L_{\rm IG}. 
\end{align}
Here, we consider, that the theory includes the Lagrangian density of the electro-magnetic field on weakly curved space-time. Therefore, we can rewrite the Eq.~(\ref{e19}) using the Eqs.~(\ref{e18}) and (\ref{e16}) up to the first order in terms of the gravitational field tensor $h_{\mu \nu}$ as
\begin{align}
\label{e20}
 \mathcal{L} &\approx L_{\rm QED, flat}  -  \dfrac{1}{2}h_{\mu \nu}(T^{\mu \nu} + F^{\mu \lambda} F_{\lambda}^{\phantom{\lambda} \nu} - \eta^{\mu \nu} L_{\rm QED, flat}) \\
 \label{e21}
 &\approx L_{\rm QED, flat} - \dfrac{1}{2} h_{\mu \nu} T^{\mu \nu}_{\rm QED}.
\end{align}
In Eq.~(\ref{e20}), we have identified, that the tensor in the product with gravitational field tensor $h_{\mu \nu}$ is the energy-momentum tensor of quantum electrodynamics on flat space-time $T^{\mu \nu}_{\rm QED}$. Therefore, the gravitational field $h_{\mu \nu}$ has a source $T^{\mu \nu}_{\rm QED}$ in analogy with the electro-magnetic radiation's  conserved four-current $J^\mu$ in flat space-time.

The Lagrangian density $\mathcal{L}$ in Eq.~(\ref{e21}) can be interpreted to describe a theory on a flat space-time $x^\mu$ involving the tensor gravitational field $h_{\mu \nu}$, the electro-magnetic field $A^\mu$ and the Dirac fields $\psi$ and $\bar{\psi}$. We have shown, that the GQM theory proposed by Adler~et.~al. extended with the dynamics of the free electro-magnetic field is equivalent to a physically meaningful flat space-time theory. Nevertheless, the flat space-time described by coordinates $x^\mu$ is strictly speaking not physical, but it is related to the physical one by a non-trivial metric tensor $g_{\mu \nu}$. At last it is worth to note, that the addition of the free electro-magnetic term to the Lagrangian density $L$ allows the theory to predict propagation of the electro-magnetic field on weakly curved space-time.

\section{Conclusion}

The energy-momentum tensor of the original GQM theory presented in Sec.~II has been shown to be asymmetric due to the lack of a term describing the intrinsic energy of the electro-magnetic field in the Lagrangian density. Thus, the energy-momentum tensor does not appear in the interaction Lagrangian density of the original GQM theory. After the extension of the Lagrangian density of the GQM theory with the free electro-magnetic field term on a weakly curved space-time, a symmetric energy-momentum tensor of flat space-time quantum electrodynamics appeared in the interaction Lagrangian density. It has been shown, that the Lagrangian density of the theory developed from the principle of general covariance~\cite{Adler2012} extended by the free electro-magnetic field term is equivalent to a flat space-time theory, where the gravitational field is treated as an ordinary tensor field, in analogy to the electro-magnetic field.

\appendix

\section{Derivation of Energy Momentum Tensor $T^{\mu \nu}_{\rm N}$ with Second Noether Theorem}

First we introduce the canonical energy-momentum tensor $\Theta^{\mu \nu}$. Then, we show which requirements are put on a general energy-momentum tensor $T^{\mu \nu}$, which can replace the canonical energy-momentum tensor $\Theta^{\mu \nu}$ in a studied theory. Finally, we derive the energy-momentum tensor $T^{\mu \nu}_{\rm N}$ from the Lagrangian density $L_{\rm QED, flat}$ in Eq.~(\ref{e8}) with the second Noether theorem and discuss, wether it still holds properties put on a generalized energy-momentum tensor.

The canonical energy-momentum tensor $\Theta^{\mu \nu}$ can be derived from the Lagrangian density $L$ in Eq.~(\ref{e1}) as follows
\begin{align}
\nonumber
 \dfrac{\partial L}{\partial x^\nu} &= \dfrac{\partial L}{\partial \psi} \dfrac{\partial \psi}{\partial x^\nu} + \dfrac{\partial L}{\partial(\partial_\mu \psi)} \dfrac{\partial^2 \psi}{\partial x^\mu \partial x^\nu} + \\
 \nonumber
 &\dfrac{\partial L}{\partial \bar{\psi}} \dfrac{\partial \bar{\psi}}{\partial x^\nu} + \dfrac{\partial L}{\partial(\partial_\mu \bar{\psi})} \dfrac{\partial^2 \bar{\psi}}{\partial x^\mu \partial x^\nu} + \\
 \label{ea1}
 & \dfrac{\partial{L}}{\partial A_\mu} \dfrac{\partial A_\mu}{ \partial x^\nu}.
\end{align}
With the utilization of the Euler-Lagrange equations for the Dirac field $\psi$
\begin{equation}
 \label{ea2}
 \dfrac{\partial L}{\partial \psi} - \dfrac{\partial }{\partial x^\mu} \left( \dfrac{\partial L}{\partial(\partial_\mu \psi)} \right) = 0
\end{equation}
and the same equation, but for the field $\bar{\psi}$, we arrive at the equation
\begin{align}
 \label{ea3}
 \dfrac{\partial L}{\partial x^\nu} = \dfrac{\partial}{\partial x^\mu} \left(\dfrac{\partial L}{\partial(\partial_\mu \psi)}(\partial_\nu \psi) +  \dfrac{\partial L}{\partial(\partial_\mu \bar{\psi})}(\partial_\nu \bar{\psi})\right) + \dfrac{\partial{L}}{\partial A_\mu} \dfrac{\partial A_\mu}{ \partial x^\nu}.
\end{align}
Equation~(\ref{ea3}) can be finally arranged into form, in which the canonical energy-momentum tensor appears
\begin{align}
\nonumber
 &\dfrac{\partial}{\partial x^\mu} \underbrace{\left(\dfrac{\partial L}{\partial(\partial_\mu \psi)}(\partial_\nu \psi) +  \dfrac{\partial L}{\partial(\partial_\mu \bar{\psi})}(\partial_\nu \bar{\psi}) - \delta^\mu_\nu L  \right)}_{\Theta^\mu_{\phantom{\mu}\nu}} = \\
 \label{ea4}
 &- \dfrac{\partial{L}}{\partial A_\mu} \dfrac{\partial A_\mu}{ \partial x^\nu}.
\end{align}

We have assumed, that except the fields $\psi$ and $\bar{\psi}$, the Lagrangian density $L$ contains an external electro-magnetic field $A^\mu$. The electro-magnetic field is defined as a function of space and time and may exert a force influencing the motion of the Dirac fields $\psi$ and $\bar{\psi}$. The force term occupies the right-hand side of Eq.~(\ref{ea4}). Therefore, in the studied case the canonical energy-momentum tensor $\Theta^{\mu \nu}$ is not conserved, but its components are influenced by the external force field.

In the general theory of relativity, only symmetric energy-momentum tensors have physical significance. They act as a source of space-time curvature. Therefore, it is desirable to symmetrize the (generally asymmetric) canonical energy-momentum tensor $\Theta^{\mu \nu}$. The canonical energy-momentum tensor $\Theta^{\mu \nu}$ defined by Eq.~(\ref{ea4}) and Lagrangian $L_{\rm flat}$ in Eq.~(\ref{e2}) is not symmetric. This occurs, because the fields $\psi$ and $\bar{\psi}$ possess intrinsic angular momentum -- spin. The symmetric energy-momentum tensors are the Hilbert and the Noether energy-momentum tensor derived by her second theorem \cite{Baker2021,Freese2022}. In this paper, we derive the energy momentum tensor by the second Noether theorem.

Energy-momentum tensor of a theory related to a particular Lagrangian density is not given uniquely. The canonical energy momentum tensor $\Theta^{\mu \nu}$ of a theory can be extended by the four-divergence of the Belinfante tensor $B^{\lambda \mu \nu}$ in order to obtain new energy momentum tensor $T^{\mu \nu}$
\begin{equation}
\label{ea5}
 T^{\mu \nu} = \Theta^{\mu \nu} + \partial_\lambda B^{\lambda \mu \nu}.
\end{equation}
It is required, that the divergence of the new energy-momentum tensor and the canonical energy-momentum tensor remain the same
\begin{equation}
\label{ea6}
 \partial_\mu T^{\mu \nu} = \partial_\mu \Theta^{\mu \nu}.
\end{equation}
This holds if the Belinfante tensor $B^{\lambda \mu \nu}$ is anti-symmetric in its first two indices
\begin{equation}
\label{ea7}
 B^{\lambda \mu \nu} = - B^{\mu \lambda \nu}.
\end{equation}
As a consequence of the anti-symmetry of the Belinfante tensor in Eq.~(\ref{ea7}), the spatial-volume integrals of $T^{0 \nu}$ and $\Theta^{0 \nu}$ are equal
\begin{align}
\label{ea8}
P^\nu &= \int d^3x\, T^{0 \nu} = \\
\label{ea81}
&= \int d^3 x\, \Theta^{0 \nu} + \int d^3 x\,\partial_0 B^{00 \nu} + \int d^3 x\,\partial_j B^{j 0 \nu}\\
\label{ea82}
&= \int d^3 x\, \Theta^{0 \nu}.
\end{align}
The second term in Eq.~(\ref{ea81}) is equal to zero since $B^{00\nu} = 0$ due to the antisymmetry of the Belinfante tensor in its first two indices. The third term is equal to zero, since the volume of the integration is assumed to expand its boundary to infinity and the terms $B^{j0\nu}$ are assumed to decrease faster than $1/r^2$. The equality of volume integrals in Eqs.~(\ref{ea8}) and (\ref{ea82}) is necessary, since both energy-momentum tensors $\Theta^{\mu \nu}$ and $T^{\mu \nu}$ have to provide the same values of four-momentum components (observables) $P^\nu$.

According to Freese~\cite{Freese2022} the second Noether theorem states, that the energy-momentum tensor $T^\mu_{\phantom{\mu}\nu,\mathrm{N}}$ is defined by the equation
\begin{align}
\label{ea9}
 T^{\mu}_{\phantom{\mu} \nu,\mathrm{N}} &= \mathscr{D}^{\mu}_{\phantom{\mu}\nu}[\psi, \bar{\psi}] - \delta^{\mu}_\nu L;\\
 \nonumber
 \mathscr{D}^\mu_{\phantom{\mu} \nu}[\psi,\bar{\psi}] & = - \dfrac{\partial}{\partial(\partial_\mu \xi^\nu)} \left[ \dfrac{\partial L}{\partial \psi} \Delta \psi + \dfrac{\partial L}{\partial(\partial_\lambda \psi)} \Delta (\partial_\lambda \psi) \right. \\
 \label{ea10}
 & \left. \dfrac{\partial L}{\partial \bar{\psi}} \Delta \bar{\psi} + \dfrac{\partial L}{\partial(\partial_\lambda \psi)} \Delta (\partial_\lambda \psi) \right].
\end{align}
$\xi^\mu$ is a component of the local translational vector, $\Delta \psi$ and $\Delta (\partial_\lambda \psi)$ are total variations of the spinor field $\psi$ and its derivative $\partial_\lambda \psi$. For formulas defining the total variations, please see \cite{Freese2022}. It is worth to note, that the quantity $\mathscr{D}^\mu_{\phantom{\mu} \nu}$ is a linear operator with respect to the Lagrangian density $L$. Thus, the calculations can be conveniently partitioned. Particularly, the energy-momentum tensor $T^\mu_{\phantom{\mu}\nu,\mathrm{N}}$ for the Lagrangian density of the free Dirac field $\psi$ and $\bar{\psi}$ on the flat space-time can be calculated first. Then, the energy-momentum tensor calculated from the interaction Lagrangian density (incorporating fields $\psi$, $\bar{\psi}$ and $A^\mu$) has to be added to the one previously obtained. The final form of the energy-momentum tensor is \cite{Markoutsakis2019}
\begin{align}
\nonumber
T^{\mu \nu}_{\rm N} &= \dfrac{1}{2}(\Theta^{\mu \nu} + \Theta^{\nu \mu} + q \bar{\psi} A^\nu \gamma^\mu \psi - q \bar{\psi} A^\mu \gamma^\nu \psi);\\
\label{ea11}
\Theta^{\mu \nu} &= \dfrac{i}{2} \left[ \bar{\psi} \gamma^\mu (\partial^\nu \psi) -  (\partial^\nu \bar{\psi}) \gamma^\mu \psi \right].
\end{align}
The obtained tensor $T^{\mu \nu}_{\rm N}$ is not symmetric. This is caused by the angular momentum transferred to the field $\psi$ by the external electro-magnetic field $A^\mu$.

During derivation of the energy-momentum tensor $T^{\mu \nu}_{\rm N}$, the Belinfante tensor can be identified with the expression
\begin{align}
 \label{ea12}
 B^{\lambda \mu \nu} &= \dfrac{i}{4} \left[ - \dfrac{\partial L}{\partial (\partial_\lambda \psi)} \sigma^{\mu \nu} \psi + \bar{\psi} \sigma^{\mu \nu} \dfrac{\partial L}{\partial(\partial_\lambda \bar{\psi})} \right]\\
 \label{ea13}
 &= \dfrac{1}{8} \bar{\psi}[\gamma^\lambda,\sigma^{\mu \nu}]_+ \psi;\,\sigma^{\mu \nu} = \dfrac{i}{2}[\gamma^\mu,\gamma^\nu].
\end{align}
From Eq.~(\ref{ea12}) follows, that the Blinfante tensor $B^{\lambda \mu \nu}$ is not influenced by an interaction Lagrangian density, which is independent of derivatives of spinors $\partial_\lambda \psi$ and $\partial_\lambda \bar{\psi}$. This is the case of the interaction Lagrangian density for the electro-magnetic field, see Eq.~(\ref{e2}). It can be easily verified, that the Belinfante tensor in Eq.~(\ref{ea13}) satisfies the requirement in Eq.~(\ref{ea7}). Therefore, the derived energy-momentum tensor $T^{\mu \nu}_{\rm N}$ in Eq.~(\ref{ea9}) is connected with the canonical energy-momentum tensor $\Theta^{\mu \nu}$ through the relation in Eq.~(\ref{ea5}). Moreover, the components of the four-momentum $P^\nu$  are the same, regardless of the utilized energy-momentum tensor. See Eqs.~(\ref{ea8}) and (\ref{ea82}).

The explicit value of the four-divergence of the canonical energy-momentum tensor $\Theta^{\mu \nu}$ and the energy momentum tensor $T^{\mu \nu}_{\rm N}$ derived by the second Noether theorem is equal to
\begin{equation}
\label{ea14}
 \partial_\mu \Theta^{\mu \nu} = \partial_\mu T^{\mu \nu}_{\rm N} = -\dfrac{\partial L}{\partial A_\mu} (\partial^\nu A_\mu) = q (\partial^\nu A_\mu) \bar{\psi} \gamma^\mu \psi
\end{equation}
in accordance with Eq.~(\ref{ea4}).

\bibliography{javurek.bib}
\end{document}